\documentclass[prd,twocolumn,groupedaddress,nofootinbib, superscriptaddress]{revtex4}
\pdfoutput=1

\usepackage{amsmath,amssymb,amscd}
\usepackage{listings}
\usepackage{dsfont}
\usepackage{slashed}
\usepackage{color}
\usepackage{ulem}

\usepackage[pdftex]{graphicx}
\usepackage{epstopdf}
\usepackage{subfigure}
\usepackage{epsfig}
\begin{document}

\rightline{}
\rightline{{\tiny Cavendish-HEP-18/09, DAMTP-2018-16}}

\title{Leptogenesis in Cosmological Relaxation with Particle Production}


\author{Minho~Son} 
\author{Fang~Ye}

 \affiliation{Department of Physics, Korea Advanced Institute of Science and Technology, \\
291 Daehak-ro, Yuseong-gu, Daejeon 34141, Republic of Korea}

\author{Tevong~You}

\affiliation{DAMTP, University of Cambridge, Wilberforce Road, Cambridge, CB3 0WA, UK; \\
Cavendish Laboratory, University of Cambridge, J.J. Thomson Avenue, Cambridge, CB3 0HE, UK}


\begin{abstract}
Cosmological relaxation of the electroweak scale is improved by using particle production to trap the relaxion. We combine leptogenesis with such a relaxion model that has no extremely small parameters or large e-foldings. Scanning happens after inflation---now allowed to be at a high scale---over a sub-Planckian relaxion field range for a cutoff scale of new physics up to $\mathcal{O}(100)$ TeV. Particle production by the relaxion also reheats the universe and generates the baryonic matter-antimatter asymmetry. We propose a realisation in which out-of-equilibrium leptons, produced by the relaxion, scatter with the thermal bath through interactions that violate CP and lepton number via higher-dimensional operators. Such a minimal effective field theory setup, with no new physics below the cutoff, naturally decouples new physics while linking leptogenesis to relaxion particle production; the baryon asymmetry of the universe can thus be intrinsically tied to a weak scale hierarchy.  
\end{abstract}

\maketitle 



\section{Introdcution}
Cosmological relaxation of the electroweak scale~\cite{GKR} has gained much interest in recent years as an alternative approach to the hierarchy problem that allows decoupled new physics without fine-tuning the Higgs mass. In the absence of new physics at the LHC, this is an increasingly motivated scenario, but the difficulty of including baryogenesis in a realistic relaxation model has been one of the main hindrance to further developments. We address this here by combining leptogenesis with recent progress towards a more viable relaxation mechanism.

The general relaxation idea is as follows. Consider an axion (the so-called relaxion), $\phi$, whose shift symmetry is softly broken by some dimensionful parameter $g$; this can arise, for example, in axion monodromy~\cite{monodromy} and clockwork constructions~\cite{clockwork}.
Below the Standard Model Effective Field Theory (SM EFT) cutoff $\Lambda$, including all interactions not forbidden by symmetries, the most general Lagrangian contains the terms,
\begin{equation}
\mathcal{L}_{\text{SMEFT}+\phi} \supset (\Lambda^2 - g\phi)|h|^2 + g\Lambda^2\phi + \text{...} \, ,
\end{equation}
where $h$ is the Higgs doublet with mass $\mu^2 \sim \Lambda^2$, and the ellipses denote higher-order terms in the soft-breaking potential $V(g\phi)$. This potential causes a slope along which $\phi$ rolls during the early universe. As it rolls, it scans an effective Higgs mass $\mu^2|_\text{eff.} \equiv \Lambda^2 - g\phi$. At negative values of $\mu^2|_\text{eff.}$, the Higgs' vacuum expectation value $v$ is non-zero. All that is then needed to explain why $v \ll \Lambda$ is for a backreaction to switch on and trap the relaxion when $\mu^2|_\text{eff.}$ is small and negative. 

In the original GKR mechanism~\cite{GKR}, the trapping backreaction acted on the relaxion's periodic potential,
\begin{equation}
V(\phi) \supset \Lambda_c^4 \cos(\phi/f_p) \, ,
\end{equation}
whose barriers $\Lambda_c^4 \simeq \Lambda_\text{QCD}^3 v$ will grow with a linear dependence on $v$ until they are sufficiently large to compensate for the slope. Unfortunately this creates several problems: if $\phi$ is the QCD axion, it no longer solves the strong CP problem without some additional mechanism, and if the periodic potential is due to the condensate of another gauge group, then it reintroduces new physics near the weak scale. Moreover, for the barrier to trap the relaxion at the weak scale requires $g \sim 10^{-31}$ GeV. Despite being technically natural (the shift symmetry is restored as $g$ goes to zero) such a tiny value leads to conflict with the weak gravity conjecture~\cite{WGC} and exponentially long e-foldings of super-Planckian scanning during inflation.

An improvement comes from trapping using particle production~\cite{HMT, TY}. 
The relaxion's shift symmetry permits an anomalous coupling to gauge bosons and a derivative coupling to fermions, 
\begin{equation}
\mathcal{L}_\phi \supset -\frac{1}{4}\frac{\alpha_V}{f_V}\phi F_{\mu\nu}\tilde{F}^{\mu\nu} + \frac{\partial_\mu\phi}{f_L}J^{5\mu} \, ,
\label{eq:relaxiongaugefermioncoupling}
\end{equation}
where $F_{\mu\nu}$ is a gauge field strength and $J^{5\mu}$ a fermionic current.
In a minimal setup which is our baseline assumption, such a fermion coupling can arise at low energies through renormalization, involving the coupling to the gauge boson $f_V$, despite its absence at high energies.
The exponential production of gauge bosons is an efficient source of friction~\cite{HMT, TY, relaxationafterreheating, relaxationSchwinger, relaxionServant, particleproduction:Anber, particleproduction:Kofman}. In the models of Refs.~\cite{HMT, TY, relaxionServant} it is an intrinsic part of the backreaction mechanism, since the periodic potential barriers no longer depend on $v$. In this case the relaxion initially has sufficient kinetic energy to roll over them. We follow the approach of Hook and Marques-Tavares (HMT)~\cite{HMT}, where the $v$-dependence of the backreaction mechanism resides in electroweak gauge boson production. After describing the essential features of the HMT model in the next Section, we then show how it can be combined in a natural way with leptogenesis during reheating.

\section{Cosmological Relaxation with Particle Production}
\label{sec:particleproduction}

\begin{figure*}
\centering
\includegraphics[width=0.435 \textwidth]{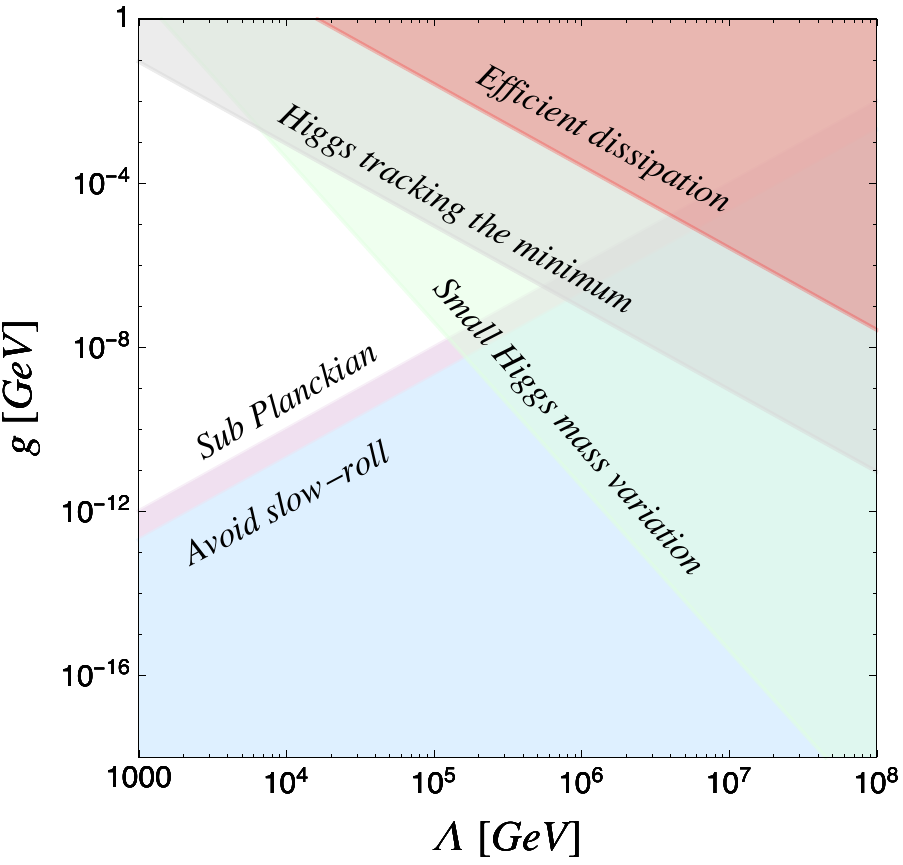} \quad \quad
\includegraphics[width=0.43 \textwidth]{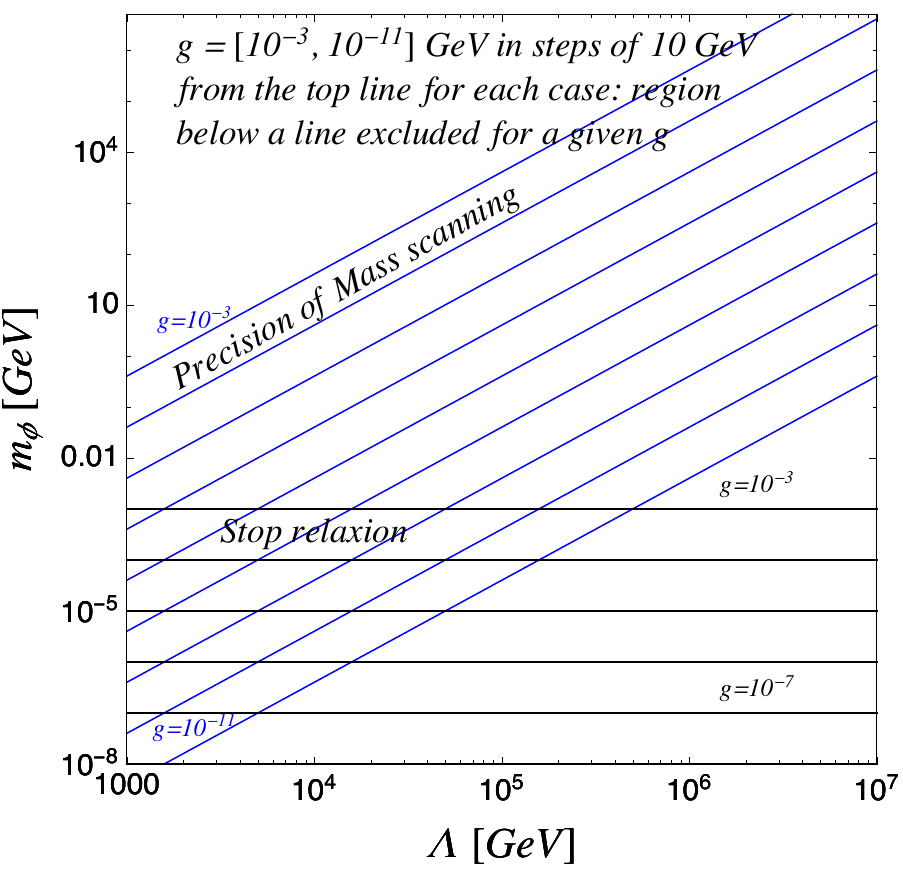} 
\caption{Theoretical constraints on cosmological relaxation with the particle production in ($\Lambda,\, g)$ plane (left) and ($\Lambda,\, m_\phi)$ plane (right). The shaded regions in the left panel are excluded. For a chosen $g$, region below a line is excluded for each type of constraint in the right panel. We set $\Lambda_c \sim \Lambda$ in both panels as it is favored according to our numerical simulation. The explanation of the constraints is given in Appendix~\ref{app:constraints}.
\label{fig:gexclusion}}
\end{figure*}

In this section we briefly review the cosmological relaxation with particle production proposed by HMT~\cite{HMT}. 
The HMT model can work either before, during, or after inflation. Relaxation during inflation requires a very low Hubble scale, $H < v$; we therefore focus on the variation in which scanning happens after inflation ends, which has the added benefit of allowing high-scale inflation.
After inflation ends, the inflaton decays to a hidden sector~\footnote{The hidden sector energy density must eventually become sub-dominant to the visible sector through a faster energy density scaling.}; this ensures the relaxion will be scanning the zero-temperature Higgs potential. The relaxion is initially displaced with an initial field value $\phi_0 > \Lambda^2 / g$ where the effective Higgs mass is negative and $v$ is large. As it scans down to smaller values, over a typical field range $\Delta\phi \sim \Lambda^2/g$, the value of $v$ decreases until the electroweak gauge bosons are light enough to be produced~\footnote{A coupling to photons must be sufficiently suppressed from the start, for example in an ultra-violet completion with $SU(2)_L \times SU(2)_R$ left-right symmetry~\cite{HMT}. 
}. This happens when $v \sim \dot{\phi}_v/f_V$, where $\dot{\phi}_v \gtrsim \dot{\phi}_0$, and determines the weak scale hierarchy $v \ll \Lambda$ in a technically natural way. Through particle production, the kinetic energy of the relaxion is converted into the temperature of the visible sector's thermal bath, $T^4 \lesssim \dot{\phi}_v^2 $, thus reheating the universe in the process.

Note that, unlike in the original GKR model, the condensate $\Lambda_c$ of the periodic potential is due to a hidden sector and does not depend on the Higgs' vacuum expectation value $v$. The relaxion must initially have enough kinetic energy to overcome the barriers, $\dot{\phi}_0^2 > \Lambda_c^4$. The initial condition for a nonzero relaxion velocity can be set in several ways after exiting high scale inflation: the inflaton preheating into hidden sector gauge bosons or fermions to which the relaxion couples could temporarily act as a background source in its equation of motion; alternatively, an inflaton-relaxion coupling $\kappa \sigma \phi$ can act as a faster effective slow-roll slope during inflation so that it exits with a velocity $\dot{\phi} \sim \kappa \sigma / H_I$. If the inflationary scale is low enough, $H_I \lesssim g\Lambda^2 / \Lambda_c^2$, then the shift-symmetry breaking slope alone gives sufficient slow-roll velocity.

The tachyonic condition for exponential gauge boson production is set by the equation of motion for the 
transverse polarization modes of
gauge boson $A_\mu \equiv \{ Z_\mu, W_\mu^\pm \}$ as
\begin{equation}
\ddot{A}_\pm + \omega_\pm^2 A_\pm = 0 \, ,
\end{equation}
where
\begin{equation}
\omega_\pm^2 = k^2 + m_A^2 \pm k\frac{\dot\phi}{f_V} \, . 
\label{eq:omega}
\end{equation}
We therefore see that only the solutions $A_{\pm}(k) \propto \text{exp}(i\omega_\pm t)$ with low momentum modes $k  < m_A$ experience exponential growth once the gauge boson mass $m_A \propto v$ drops below the dissipation threshold $\sim \dot{\phi}_v/f_V$. 
We neglected Hubble here, since the dissipation and trapping will happen on a shorter timescale, and assumed zero temperature. 
 In a plasma at finite temperature $T$, Eq.~\ref{eq:omega} can be shown to be approximately given by~\cite{HMT}
\begin{equation}
\omega^2 - k^2 \mp k\frac{\dot{\phi}}{f_V} \simeq \frac{T^2 |\omega|}{k} \, .
\end{equation}
In this case there can always be exponential production for  $k \sim \dot{\phi}/f_V$ with $-i \omega \sim \frac{(\dot{\phi}/f_V)^3}{T^2}$~\footnote{For non-Abelian gauge bosons the plasma induces a magnetic mass that restricts the possibility of a tachyonic solution, so only the $U(1)_Y$ Abelian gauge boson will be exponentially produced at finite temperature. }. We emphasize here that although the particle production starts to occur in a zero-temperature background, the produced gauge bosons quickly create a thermal bath via electroweak interactions. Note that the amount of energy density transferred from the relaxion to the tachyonic gauge bosons after $\mathcal O(1)$ field excursion is $\Delta V\sim \Lambda^4$. Those gauge bosons, whose center of mass energy is at $\mathcal O(\Lambda)$, thermalize with the Standard Model particles with interaction rates $
\Gamma\sim\alpha_{EW}^2\Lambda$ much greater than Hubble.
Since the end of relaxation will occur in a thermal bath, the inverse of the tachyonic plasma frequency gives the relevant timescale $\tau$ for particle production to lose enough energy to reach the trapping threshold at $\dot{\phi}_c \sim \Lambda_c^2$,
\begin{equation}
\tau \sim \frac{T^2 \dot{\phi}_v^3}{\Lambda_c^6 v^3} \, ,
\end{equation}
where we substituted $f_V \sim \dot{\phi}_v / v$. This timescale is taken to be faster than Hubble, $\tau \lesssim 1/H$ (note that this is not Hubble during inflation but when the relaxion is able to scan the entire field range $\Delta\phi \sim \Lambda^2 / g$ at  $H \sim g$). Moreover, the relaxion must not roll past the Higgs mass scale before being trapped, $\int g\dot{\phi}dt < v^2$, which leads to the constraint
\begin{equation}
 \frac{g T^2 \dot{\phi}_v^3}{\Lambda_c^4} < v^5 \, .
 \label{eq:overshoothiggsmass}
\end{equation}
For $g \sim \Lambda^2 / M_p$ (the value that saturates the bound of the sub-Planckian field range requirement $\Delta\phi \lesssim M_p$) Eq.~\ref{eq:overshoothiggsmass} places an upper limit on the cutoff $\Lambda$, 
\begin{equation}
\Lambda \lesssim \left(\frac{M_p v^5 \Lambda_c^4}{T^2 \dot{\phi}_v^3} \right)^\frac{1}{2} \, .
\label{eq:lambdacutoff}
\end{equation}
This can be maximised for $\dot{\phi}_v \sim \Lambda_c^2 \sim T^2$. However, a stronger bound comes from requiring the relaxion energy density to be sub-dominant during scanning, $H^2 \gtrsim V(\phi) / M_p^2 \Rightarrow \Lambda \lesssim \sqrt{H M_p}$.
A conservative bound can also be placed assuming $\dot{\phi}_v \sim T^2 \sim \Lambda^2$ in Eq.~\ref{eq:lambdacutoff} such that $\Lambda \lesssim (M_p v^5)^\frac{1}{6} \sim 10^5$ GeV. To maximise the reheating temperature for leptogenesis we shall take this as our typical upper limit for a sub-Planckian field range.

The decay constant $f_p$ must also allow for multiple minima, 
\begin{equation}
\frac{\Lambda_c^4}{f_p} \gtrsim g \Lambda^2,
\end{equation}
 each separated by less than the weak scale,
 \begin{equation}
 g f_p < v^2.
 \end{equation} 
The couplings in Eq.~\ref{eq:relaxiongaugefermioncoupling} can induce the irreducible coupling to photons through loops at low energies~\cite{Bauer:2017ris, photophobicALP}, 
\begin{equation}\label{eq:fgamma:IR}
\frac{1}{f_\gamma} = \frac{2\alpha}{\pi \sin ^2 \theta_w f_V} B_2 \left (\frac{4m_W^2}{m_\phi^2} \right )
+ \sum_F \frac{N_c^F Q_F^2}{2 \pi^2 f_F} B_1 \left ( \frac{4m^2_F}{m^2_\phi} \right )~.
\end{equation}
The fermion couplings $f_F$ in Eq.~\ref{eq:fgamma:IR} is given by 
\begin{equation}
\frac{1}{f_F} =
- \frac{3\alpha^2}{4 f_V} \Big [ \frac{3}{4 \sin^4 \theta_w} - \frac{1}{\cos^4\theta_w} \left (Y_{F_L}^2 + Y_{F_R}^2 \right ) \Big ] {\rm log} \frac{\Lambda^2}{m^2_W}~,
\end{equation}
where $Y_{F_{L,\, R}}$ denote the hypercharges of the left and right handed fermions.
The functions $B_{1,\, 2}$ in Eq.~\ref{eq:fgamma:IR} is defined as
\begin{equation}
\begin{split}
 B_1(x) = 1 - x [f(x)]^2,\quad
 B_2(x) = 1 - (x-1)[f(x)]^2~,
\end{split}
\end{equation}
where
\begin{eqnarray}
f(x) & \sim &
\left\{ 
\begin{array} {ll}
\arcsin \frac{1}{\sqrt{x}} & x \geq 1\\
\frac{\pi}{2} +\frac{i}{2}\log \frac{1+\sqrt{1-x}}{1-\sqrt{1-x}} & x<1~.
\end{array}
\right.
\end{eqnarray}
The induced photon coupling results in the tachyonic production of massless photons which can spoil the mechanism. It can be suppressed as long as the timescale for the photon production is larger than the Hubble time~\cite{relaxionServant},
\begin{equation}\label{eq:fadissipate}
t_\gamma\sim T^2f_\gamma^3/\dot{\phi}^3>H^{-1}.
\end{equation}
While the irreducible coupling to photons is also subject to the astrophysical and phenomenological constraints, those constraints are weak for the relaxion mass range of interest, namely $m_\phi \gtrsim \mathcal{O}({\rm GeV})$.

\begin{figure}[h]
\begin{center}
\includegraphics[width=0.47 \textwidth]{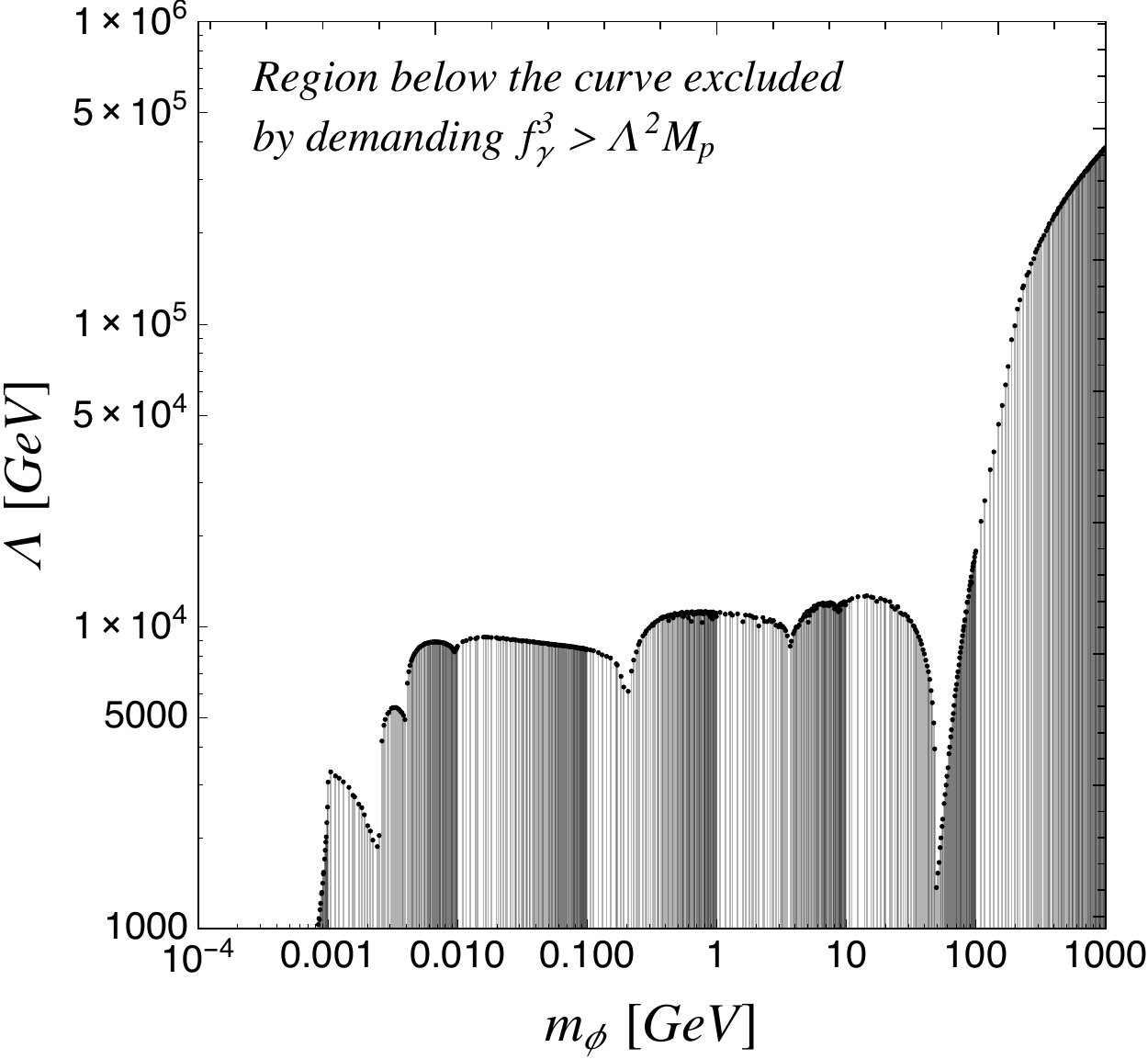}
\end{center}
\caption{Bounds on $\Lambda$ as a function of $m_\phi$ from requiring dissipation via photons to be sub-dominant. The shaded region below the curves are excluded. The constraint on the legend is derived from Eq.~\ref{eq:fadissipate}.
\label{fig:fgammaexclusion}}
\end{figure}

Throughout this paper we will impose all constraints for successful relaxation with particle production to take place, similarly to Ref.~\cite{relaxionServant}. The allowed parameter space for $g$ and $m_\phi$ as a function of $\Lambda$ are shown in Fig.~\ref{fig:gexclusion}. The constraints coming from not allowing dissipation via massless photon production to dominate are displayed in Fig.~\ref{fig:fgammaexclusion}.

\section{Leptogenesis}

We now turn to the task of implementing leptogenesis from cosmological relaxation with particle production. The relatively low temperatures achievable in the HMT model restrict the possible scenarios if we wish to avoid reintroducing new physics below the SM EFT cutoff. Reheating to the threshold of some new, heavy particle whose out-of-equilibrium decay is responsible for generating the baryon asymmetry requires reheating above the cutoff or adding new physics below it. Other baryogenesis approaches are also severely restricted by the particular requirements of the relaxion mechanism. Here, we instead make use of leptogenesis generated by inelastic scattering between leptons from the relaxion and leptons in the thermal bath~\cite{Baldes, Hamada}. All three of Sakharov's conditions are satisfied---the leptons produced by the relaxion are out-of-equilibrium, and scattering proceeds via higher-dimensional operators that violate lepton number, with CP-violating interactions. The resulting lepton asymmetry number density $n_L$ will then be converted to a baryon asymmetry $n_B$ by the electroweak sphaleron process, 
\begin{equation}\label{eq:nBtos:nLtos}
\frac{n_B}{s} \simeq \frac{28}{79}\frac{n_L}{s} \sim \mathcal{O}(10^{-10}) \, .
\end{equation} 
The asymmetry is normalised to the entropy density $s = (2\pi^2 / 45) g_* T^3$, where $g_* \sim 10^2$ is the number of relativistic degrees of freedom. An order of magnitude estimate of the baryon asymmetry is sufficient for our purpose, as we typically neglected $\mathcal{O}(1)$ factors in our relaxion estimates. 

Remarkably, all of the ingredients for this mechanism are already present in the relaxion setup. The 
loop-induced 
derivative lepton current coupling in Eq.~\ref{eq:relaxiongaugefermioncoupling} 
in our minimal setup 
 not only respects the shift symmetry, but was previously necessary to allow the thermal abundance of the relaxion to decay away below the decoupling temperature. Also, operators of higher mass dimension suppressed by the scale of new physics are generically expected to be present in a low energy effective theory. The most general effective Lagrangian for the SM EFT can be written as
\begin{equation}
\mathcal{L}_\text{SMEFT} \supset \mathcal{L}_\text{SM} + \frac{c^{(5)}}{\Lambda_5}\mathcal{O}^{(5)} + \sum_i\frac{c^{(6)}_i}{\Lambda^2_{6, i}}\mathcal{O}_i^{(6)} + \sum_i\frac{c^{(7)}_i}{\Lambda^3_{7, i}}\mathcal{O}_i^{(7)} + \text{...} \, ,
\end{equation}
where $c_i$'s are Wilson coefficients which we would take to be order one.

The unique dimension 5 operator is the Weinberg operator, 
\begin{equation}
\mathcal{O}^{(5)} = Lh Lh\, ,
\end{equation}
where we kept $SU(2)_L$ and flavour indices implicit, $h$ is the Higgs doublet field and $L$ the left-handed lepton doublet. It generates a Majorana mass for neutrinos when the Higgs field gets a vacuum expectation value. The light neutrino mass bound $m_\nu \sim 0.1$ eV implies that the corresponding operator scale is $\Lambda_5 \sim 10^{14}$ GeV for an $\mathcal{O}(1)$ Wilson coefficient $c^{(5)}$. The contribution of this operator to the baryon asymmetry is then typically negligible for the low reheating temperatures of the relaxion, and thus we focus on dimension-7, lepton-number-violating operators generated at a scale $\Lambda_7$. As an illustrative example we shall take the operator
\begin{equation}
\mathcal{O}^{(7)} = L h \bar{e}^c \bar{u}^c d^c \, .
\label{eq:dim7op1}
\end{equation}
In the notation of Ref.~\cite{Deppisch}, the field $e$ is a right-handed lepton and $u, d$ are up- and down-type right-handed quarks, respectively, and the corresponding coefficient $c^{(7)}_{ab}$ contains indices $a, b$ representing the flavours of $L$ and $e$ respectively. Note that there is a lower bound on the scale $\Lambda_7$ coming from the contribution of dimension-7 operators to the neutrino mass, as studied for example in Refs.~\cite{DelAguila, Deppisch}. For the operator of Eq.~\ref{eq:dim7op1} this bound is low enough to be negligible; other operators have stricter bounds, 
and some of them will be discussed in below (see discussion around Eq.~\ref{O7:another}).

\begin{figure}[h]
\begin{center}
\includegraphics[width=0.2 \textwidth]{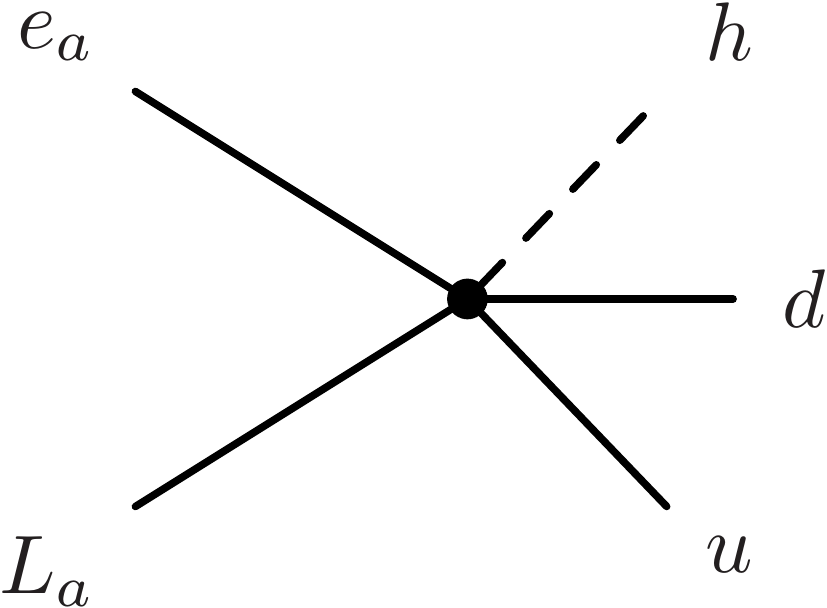}
\includegraphics[width=0.2 \textwidth]{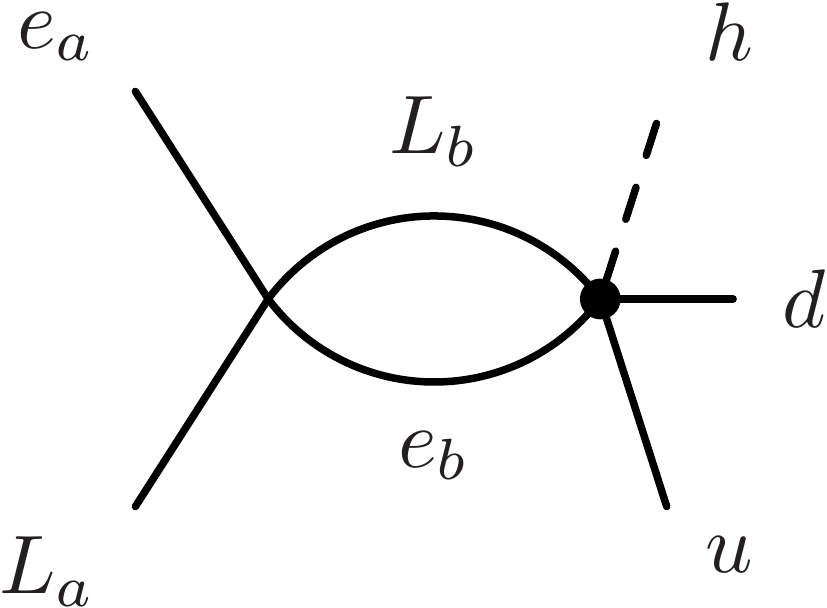}
\end{center}
\caption{Tree-level and one-loop Feynman diagrams involving the lepton-number-violating dimension-7 operator (\ref{eq:dim7op1}) and the four-fermion operator (\ref{eq:dim6op}), whose interference violates CP. 
\label{fig:feyn1}}
\end{figure}

We also have a four-fermion dimension-6 operator, 
\begin{equation}
\mathcal{O}^{(6)} = \left(\bar{L}_a\gamma^\mu L_b \right)\left(\bar{e}_c \gamma_\mu e_d \right) \, ,
\label{eq:dim6op}
\end{equation}
with complex coefficients $c^{(6)}_{abcd}$ and a scale $\Lambda_6$, whose contribution to the one-loop diagram is responsible for CP violation in the interference term with the tree-level diagram~\footnote{An asymmetry is generated despite being at leading order in the lepton-number-violating coupling; this does not contradict the Nanopoulos-Weinberg theorem~\cite{NanopoulosWeinberg} which assumes all interactions violate lepton or baryon number, as discussed in Refs.~\cite{NanopoulosWeinberg1, NanopoulosWeinberg2, NanopoulosWeinberg3}.}, shown in Fig.~\ref{fig:feyn1}. The labels $a,b,c,d$ are flavour indices. 
The efficiency factor $\epsilon$ for the asymmetry in the scattering can be parametrised as the difference between the interaction rate of the processes $\bar{L} e \to \bar{h} u \bar{d}$ and $L \bar{e} \to h \bar{u} d$,  
\begin{equation}
\epsilon_{a} = \frac{\sigma(\bar{L}_a e_a \to \bar{h} u \bar{d}) - \sigma(L_a \bar{e}_a \to h \bar{u} d)}{\sigma(\bar{L}_a e_a \to \bar{h} u \bar{d}) + \sigma(L_a \bar{e}_a \to h \bar{u} d)} \, .
\label{eq:epsilondef}
\end{equation}
Since the phase space factor cancels in the ratio, we only need to evaluate $\sigma \propto |\mathcal{M}|^2$ where the amplitude $\mathcal{M}$ can be written as 
\begin{equation}
\mathcal{M}(L_a \bar{e}_a \to h \bar{u} d) \propto \frac{c_{a}^{(7)}}{\Lambda_7} + \sum_{b} \frac{c_{b}^{(7)}}{\Lambda_7}\frac{2 c_{ab}^{(6)}}{\Lambda_6} I \, .
\end{equation}
We defined $c^{(7)}_{aa} \equiv c^{(7)}_a$ and $c^{(6)}_{abab} \equiv c^{(6)}_{ab}$, where the lepton fields are diagonalised in $\mathcal{O}^{(7)}$ with real coefficients, while $c^{(6)}$ is complex in general. $I$ is a loop factor that must also have an imaginary part from on-shell particles running in the loop for Eq.~\ref{eq:epsilondef} to be non-zero, since it evaluates to  
\begin{equation}\label{eq:epsilona}
\epsilon_a \simeq  \frac{4}{\Lambda_6^2}\text{Im}I \sum_b \frac{c_b^{(7)}\text{Im}c_{ab}^{(6)}}{c_a^{(7)}} \, ,
\end{equation}
where $\text{Im}I = p^2 / (8\pi)$, and $p^2$ is the square of the four-momentum sum of the initial state leptons. 
When a lepton pair is just produced from the perturbative decay of the relaxion with mass $m_\phi$
through the induced coupling in Eq.~\ref{eq:relaxiongaugefermioncoupling}, a typical energy of the lepton is $\sim m_\phi$. These leptons then scatter with the thermal bath via electroweak interactions, and eventually become thermal. The thermalization time $t_L\sim \Gamma^{-1}\sim 1/T\sim 1/\Lambda$ is short, compared to the Hubble time, but would be long enough for the out-of-equilibrium  leptons to generate the observed small lepton asymmetry through the lepton-number-violating processes.
Before reaching a thermal distribution
, the out-of-equilibrium leptons will have energies distributed from $m_\phi/2$ to $3T$ as they are upscattered by the thermal bath, corresponding to $6 m_\phi T \lesssim p^2 \lesssim (6T)^2$~\footnote{
While the exact evaluation of the interaction rates, here and in the following, should take into account all the thermal initial states, we find that extrapolating the zero-temperature results by replacing $p^2$ with $6mT$ or $(6T)^2$ serves a good approximation within a factor of $2$.
}. Since the dominant contribution to the asymmetry will come from the higher energy non-thermal leptons we mainly use $p_\text{max}^2 \simeq (6T)^2$ as an upper bound in our estimates, while also 
examining 
 the $p_\text{min}^2 \simeq 6 m_\phi T$ points as a conservative lower bound case. 

In a standard approximate picture for the perturbative decays of the relaxion, since $m_\phi > H$ at the time of trapping, we may treat the oscillations of $\phi$ in its local minimum as a gas of non-relativistic particles whose equation of state is that of matter. Its number density is given by
%
\begin{equation}
n_\phi (t) \sim \rho_\phi (t)/ m_\phi \, ,
\label{eq:nphimax}
\end{equation}
with the initial value for the relaxion energy density $\rho_{\phi}(0)\sim \Lambda^4$
, and the perturbative decay rate into leptons is $\Gamma_D \sim m_\phi^3/f_L^2$. This decay rate is sub-dominant to the rate of condensate scattering with the thermal bath, which goes as $\Gamma_S \sim T^2 m_\phi / f_{L,V}^2$ (we included an additional $m_\phi / T$ suppression in the naive rate to account for bose-enhancement~\cite{HMT}), and so the available number density for producing out-of-equilibrium leptons in perturbative decays is 
\begin{equation}
n_\phi^\text{min} \simeq \frac{\Gamma_D}{\Gamma_S} n_\phi \simeq \left(\frac{m_\phi}{T}\right)^2 n_\phi \, .
\label{eq:nphimin}
\end{equation}
However, this approximation does not account for the full production of fermionic modes~\cite{fermionproduction}, similarly to the case of bosons. Though the occupation number of fermions cannot be exponentially enhanced due to Pauli blocking, the generation of fermionic modes from a rolling scalar field has also been shown to give large effects~\cite{fermionproduction}. We therefore expect this to give a large contribution of non-thermal leptons whose typical energy will be of order $\dot{\phi}_v / f_L \gtrsim v$. Since a full numerical investigation of fermionic preheating-like process including backreaction and thermal effects is beyond the scope of this work, we simply account for this extra contribution by allowing the effective number density of the condensate that is available for out-of-equilibrium leptons, $n_\phi^\prime$, to vary between the minimum perturbative contribution of Eq.~\ref{eq:nphimin} and the total condensate number density Eq.~\ref{eq:nphimax},
\begin{equation}\label{eta}
\left(\frac{m_\phi}{T}\right)^2 \lesssim \frac{n_\phi^\prime}{n_\phi} \lesssim 1 \, .
\end{equation}

The number densities of the non-thermal lepton species $l_a$, of the net lepton asymmetry and the radiation energy density evolve following the Boltzmann equations. We derive them in a similar way as in~\cite{Hamada}, ignoring terms involving the Hubble parameter since the thermalization rate $\Gamma_{th}$ is much greater than the Hubble scale. These equations can be schematically written as
\begin{eqnarray}\label{eq:nli}
\dot{n}_{l_a}&=&\Gamma_D \,n_\phi^\prime\, \mathcal{B}_a-n_{l_a}\Gamma_{th},\\
\label{eq:rhoR}
\dot{\rho}_R&=&\Gamma_S \,\rho_\phi+\rho_{l_a}\,\Gamma_{th},\\
\label{eq:nL}
\dot{n}_{L}&=&4\sum_a\left[n_{l_a}{\Gamma_\text{1 LNV}}_a\, \epsilon_{1\, a}+\frac{1}{2}n_{l_a}{\Gamma_\text{2 LNV}}_{a}\, \epsilon_{2\, a}\right]\notag\\
&-&\Gamma_{wash}\, n_L,
\end{eqnarray}
where $\mathcal B_a$ is the branching fraction of the relaxion perturbatively decaying into lepton species $a$, $n_{l_a}$ denotes the number density of out-of-equilibrium leptons, $\rho_R=\frac{\pi^2}{30}g_\star T^4$ is the radiation energy density, and $n_L$ is the net lepton number density. We have used the effective number density $n_\phi^\prime$ to capture all the sources that produce out-of-equilibrium leptons. Then the first term on the right hand side (r.h.s.) of Eq.~\ref{eq:nli} represents all these contributions, normalized by the factor $\Gamma_D\mathcal B_a$, while the second term comes from the fact that by scattering with the plasma the out-of-equilibrium leptons eventually approach thermal equilibrium. 
The first term on the r.h.s. of Eq.~\ref{eq:rhoR} denotes the contribution to the radiation energy density from relaxion condensate scattering with the plasma, and the second term corresponds to the energy added to radiation by out-of-equilibrium leptons interacting with the bath.  The first and the second terms on the r.h.s. of Eq.~\ref{eq:nL} correspond to the lepton-number-violating interactions between a thermal and an out-of-equilibrium lepton, between two out-of-equilibrium leptons, respectively. The subscript 1, 2 in ${\Gamma_\text{1,2 LNV}}_a$ and $\epsilon_\text{1,2 a}$ represents the number of out-of-equilibrium leptons in an interaction, and we will drop it unless it is necessary. The last term of Eq.~\ref{eq:nL} accounts for possible washout effects that could erase the produced lepton asymmetry. 

The exact solution to the Boltzmann equations is quite complicated and beyond the scope of this work. In this work, we provide an approximate solution to the Boltzmann equations near the end of leptogenesis which is sufficient for our purpose to show how the generated lepton asymmetry evolves and how the washout interactions affect the asymmetry. The detailed procedure can be found in Appendix~\ref{app:boltzmann}, and we simply take the final result here.
The number density of the net lepton asymmetry can now be estimated as the fraction of the number density $n_\phi^\prime$ converted into pairs of out-of-equilibrium leptons with a branching ratio $\mathcal{B}$ that undergo lepton-number-violating inelastic scattering at a rate $\Gamma_{LNV}$, relative to the thermal elastic scattering rate $\Gamma_\text{th.}$, with an efficiency $\epsilon$:
\begin{equation}\label{eq:nLtos}
\frac{n_L}{s} \simeq \frac{n_\phi^\prime}{s} \sum_a 4\epsilon_{a}\, \mathcal{B}_a   \frac{{\Gamma_\text{LNV}}_a}{\Gamma_\text{th.}} \, ,
\end{equation}
where we take into account only the scattering between an out-of-equilibrium lepton and a thermal lepton. The lepton-number-violating and thermal scattering rates in Eq.~\ref{eq:nLtos} are given by 
\begin{equation}
{\Gamma_\text{LNV}}_a \simeq \frac{1}{128\pi^5}\, p^4 \left(\frac{c^{(7)}_a}{\Lambda_7^3}\right)^2 T^3 \quad \, , \quad \Gamma_\text{th.} \simeq \alpha_2 T  \, ,
\label{eq:Gamma}
\end{equation}
where $\alpha_2(\sqrt{p^2}=10^5 \text{ GeV}) \sim 0.03$ is the $SU(2)_L$ structure constant, and we assumed thermal particles follow the Boltzmann distribution. The expression in Eq.~\ref{eq:nLtos} approximates within an order of magnitude the numerical results of solving the Boltzmann equations, provided that washout effects are negligible 
\begin{equation}\label{weakwashout}
\Gamma_{wash}<H,
\end{equation}
where the rates of the washout processes such as $\bar{h}u\bar{d}\rightarrow \bar{l}e$ and $u\bar{d}\rightarrow \bar{l}eh$ etc. are estimated to be
\begin{equation}
   \Gamma_{wash} = \frac{9}{\pi^5} T^7 \left ( \frac{c^{(7)}}{\Lambda_7^3} \right )^2~.
\end{equation}
It is obvious to see that once Eq.~\ref{weakwashout} is satisfied at the onset of washout, it continues to be satisfied during the cooling period, since $\Gamma_{wash}$ decreases faster than the Hubble parameter. 
A larger $\Lambda_7$ than the cutoff $\Lambda$ is favored to suppress the washout rates for $T\sim \Lambda$, and at the same time, it becomes challenging to generate enough lepton asymmetry. As a result, only the benchmark scenario for $p^2_{\rm max}$ and the maximum number density $n_\phi^\prime = n_\phi$ successfully generates sufficient lepton asymmetry. There could be a similar-sized contribution from the scattering between two out-of-equilibrium leptons for this benchmark scenario. However, our main result would remain the same. 

\begin{figure}[h]
\begin{center}
\includegraphics[width=0.46 \textwidth]{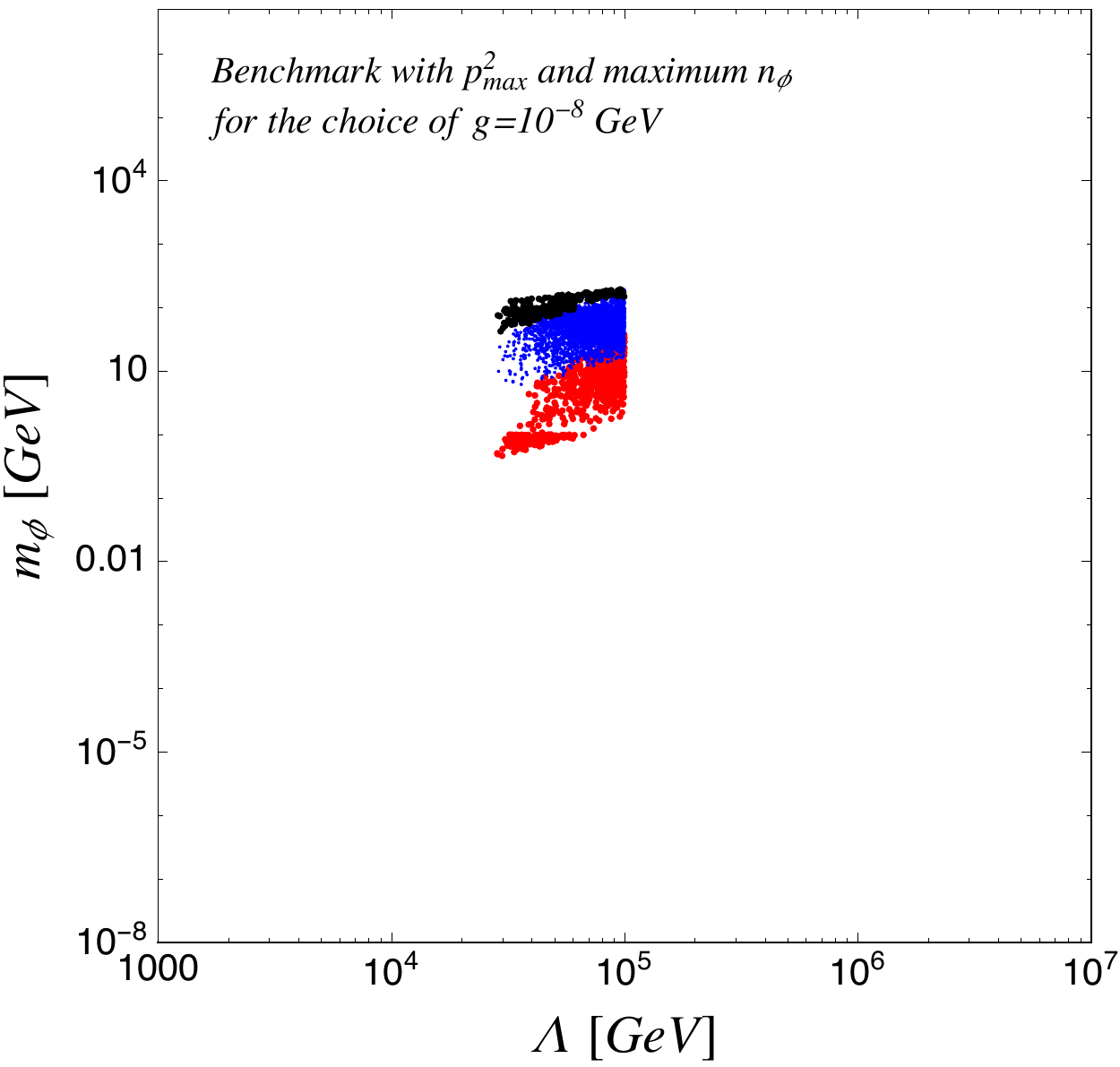}
\end{center}
\caption{Benchmark parameter space points in the $m_\phi$ vs $\Lambda$ plane that pass all constraints, for $g =10^{-8}$ GeV in the $p^2_\text{max}$, maximum $n_\phi$ scenario described in the text. The red, blue and black points are for $\mathcal{B} = 10^{-2}, 10^{-1}, 1$, respectively. 
\label{fig:benchmarks}}
\end{figure}

We set $\Lambda \sim \Lambda_{c, 6} \sim T$ for simplicity, though one should bear in mind that they only appear to be equal within an order of magnitude and can be varied independently.  We also have chosen $g \sim 10^{-8}$ GeV to push the cutoff scale up to $\Lambda\sim 10^5$ GeV while satisfying all the constraints in order for the relaxation mechanism with particle production to work~\cite{relaxionServant}, as discussed in Section~\ref{sec:particleproduction}.   We learn through our numerical simulation that the scale $\Lambda_7$ takes values of order $\mathcal{O}(10^7)$ GeV for $\Lambda \sim 10^5$ GeV, and two benchmark points are selectively shown in Table~\ref{tab:benchmark} for the purpose of illustration. For instance, for $\mathcal{B} = 1$, we can obtain the following baryon asymmetry,

\begin{align}
\frac{n_B}{s} &\sim 1.\times 10^{-10} \left(\frac{\mathcal{B}}{1}\right)  \left(\frac{T}{1.\times 10^5 \text{ GeV}}\right)^8 \left(\frac{n_\phi/s}{7.2\times 10^2}\right)  \nonumber \\
&\times \left( \frac{1.34\times10^{7} \text{ GeV}}{\Lambda_7} \right)^6 \left( \frac{1.\times 10^{5} \text{ GeV}}{\Lambda_6} \right)^2 \, .
\end{align}
A numerical scan of more allowed parameter space points for $g = 10^{-8}$ GeV, in the above scenario, is plotted in Fig.~\ref{fig:benchmarks}. The red, blue and black colour coding represents three different branching ratios, $\mathcal{B} = 10^{-2}, 10^{-1}, 1$, respectively.

We may also consider other lepton-number-violating dimension-7 operators; qualitatively, the mechanism is not much affected by the details of the specific operator, though the parameter space will be quantitatively different depending on the phase space factor. Phenomenological constraints also vary for each operator, as studied e.g. in Refs.~\cite{DelAguila, Deppisch}. The particular choice may also be theoretically motivated; for example, Ref.~\cite{DelAguila} showed that for lepton-number-violating operators with two leptons and no quarks there is a unique operator corresponding to the specific chirality of the lepton pair which are of dimension 5, 7 and 9 for $LL$, $Le$ and $ee$, respectively. The dimension-7 operator in this case is 
\begin{equation}\label{O7:another}
\mathcal{O}^{(7)} = \bar{L}\gamma^\mu e^c h (h^\dagger D_\mu \tilde{h}) \, .
\end{equation}
The lifetime of the inverse neutrinoless double-beta $(0\nu\beta\beta)$ decay is $T_{1/2} > 1.9\times 10^{25}$ years which sets a lower bound on the scale $\Lambda_7 \gtrsim 10^5$ GeV for $c^{(7)} \sim \mathcal{O}(1)$~\cite{DelAguila}. However a stronger bound comes from the operator's one-loop contribution to the neutrino mass, $m_\nu \sim \frac{v}{8\sqrt{2}\pi^2\Lambda}\,m_e\,\lambda_{7,\,ee}$, which requires $\Lambda_7 \gtrsim 10^7$ GeV. Here, the phase space factor in Eq.~\ref{eq:Gamma} also gives a larger suppression as it involves an additional particle in the final state. 
We suspect that the lowest allowed scale $\Lambda_7 \sim 10^7$ GeV might not be compatible with the sufficient lepton asymmetry although it would more easily be able to suppress the washout effect. 

\begin{table}
\centering
\begin{tabular}{|c|c|c|c|c|c|c|c|}
\hline
 $\Lambda, \Lambda_c, \Lambda_6, T$ & $\Lambda_7$ & $f_p$& $m_\phi$ & $f_V$ &  $g$ & $\mathcal{B}$\\
 \hline 
$10^5$&$1.34 \times 10^7$&$5. \times 10^{7}$ & $200$ & $5.5 \times 10^7$ & $10^{-8}$ &  1 \\
 \hline
$10^5$&$1.03 \times 10^7$&$1. \times 10^{9}$ & $10$ & $5.5 \times 10^{7}$ & $10^{-8}$ & $10^{-2}$\\
\hline
\end{tabular} 
\caption{A benchmark point in GeV (except the last column) for our relaxion leptogenesis mechanism.
}
\label{tab:benchmark}
\end{table}



\section{Conclusion}
 We proposed a model of cosmological relaxation of the weak scale with particle production that generates the baryonic matter-antimatter asymmetry while reheating the universe after inflation. For an SM EFT cutoff up to $\mathcal{O}(100)$ TeV, the model has several desirable features: it allows for high-scale inflation, scanning with a sub-Planckian field range, has no extremely small parameters, introduces no new physics below the cutoff, and achieves leptogenesis at low temperatures.

The model makes use of the thermal bath from bosons produced by the relaxion, and the lepton coupling that was already included to dilute the relaxion's thermal abundance. This necessarily leads to the production of leptons by relaxion rolling as well as through perturbative decays in the misalignment mechanism. These leptons are out-of-equilibrium and scatter with the thermal bath. The scattering will generally involve higher-dimensional operators that violate lepton number, whose effect can be large enough to generate the observed baryon asymmetry of the universe if the scale of these operators are sufficiently close to the cutoff.

From a conceptual point of view, leptogenesis in relaxation combines two approaches to understanding the smallness of the weak scale: a ``dynamical selection'' mechanism and a ``censorship'' approach~\cite{Giudice}. The relaxion mechanism ensures its evolution naturally selects a minimum with $v \ll \Lambda$, whereas tying its particle production backreaction to reheating and leptogenesis gives a cosmological censorship criteria for us living in the corner of the universe where the relaxion happened to have the right initial conditions for sufficient scanning---if it did not, the universe would be empty.  

 The lack of new physics at the weak scale that was expected to solve the hierarchy problem may mean such a solution is simply postponed to higher energies. The issue has certainly not gone away---on the contrary, it is exacerbated by the experimental null results. It is therefore worthwhile to explore alternative ways of naturally obtaining a hierarchy, with much still to be learned from dynamics in the early universe where scalar fields and Higgs-dependent phenomena can play a major r\^{o}le.

\section*{Acknowledgements}

We thank Mohamed M. Anber, Kohei Kamada, Jose Miguel No, and Michael Ramsey-Musolf for useful discussions, and the Hong Kong Institute for Advanced Study, where part of this work was carried out, for kind hospitality. TY was supported by a Junior Research Fellowship from Gonville and Caius College and partially supported by STFC consolidated grant ST/P000681/1. FY and MS were supported by Samsung Science and Technology Foundation under Project Number SSTF-BA1602-04.

\appendix
\section{Theoretical constraints for cosmological relaxation from particle production}
\label{app:constraints}
Here we list the constraints used in Fig.~\ref{fig:gexclusion}. They are similar to those in~\cite{relaxionServant} except that our $g$ is a dimensionful parameter.
\begin{enumerate}
\item Avoid slow-roll:\\
The relaxion-driven inflation should be short enough~\cite{relaxionServant},
\begin{equation}
 g \gtrsim 0.2 \frac{\Lambda^2}{M_p}~.
\end{equation}
\item Efficient dissipation:\\
The kinetic energy gaining while rolling down the potential, $\Delta K_{rolling}$, should be smaller than the amount of energy lost via the particle production, $\Delta K_{pp}$,
\begin{equation}
 -\frac{dV}{dt}\Delta t_{pp} \sim \Delta K_{rolling} \lesssim \Delta K_{pp}\sim \frac{\dot\phi^2}{2}~,
\end{equation}
where $\Delta t_{pp} \sim (9\pi^2/16)\, g^2_{EW} T^2f^3/\dot{\phi}^3$ is the time duration for the particle production.
\item Higgs tracking the minimum:\\
The scanning of the relaxion should be done while the Higgs field sits on its mininum,
\begin{equation}
 g \lesssim \frac{(v\sqrt{\lambda})^3}{\Lambda^2}~.
\end{equation}
\item Small Higgs mass variation:\\
The relaxion should not overshoot the correct Higgs mass during the time it takes to lose all of its kinetic energy,
\begin{equation}
  \Delta m_h \sim \frac{\Delta m_h^2}{m_h} \sim \frac{g}{m_h} \Delta t_{pp}\, \dot{\phi} \lesssim m_h~.
\end{equation}
\item Sub-Planckian:\\
The field excursion should be sub-Planckian,
\begin{equation}
  \Delta \phi \sim \frac{\Lambda^2}{g} \lesssim M_p~.
\end{equation}
\item Precision of mass scanning:\\
The Higgs mass must be scanned with the enough precision,
\begin{equation}
  \Delta m^2_h \sim g\, \Delta \phi \sim g\, 2\pi f \lesssim m_h^2~.
\end{equation}
\item Stop relaxion:\\
The slope of the linear potential should be smaller than that of cosine potential,
\begin{equation}
  g\Lambda^2 \lesssim \frac{\Lambda_c^4}{f}~.
\end{equation}
\end{enumerate}

\section{Approximate solution to Boltzman equations}
\label{app:boltzmann}
Here we provide the approximate solution to the Boltzmann equations in Eqs.~\ref{eq:nli},~\ref{eq:rhoR}, and~\ref{eq:nL} near the end of reheating where we assume that the universe is radiation-dominated.
For the purpose of illustration, we first consider out-of-equilibrium leptons purely from the perturbative decay of relaxions, that is, $n_\phi^\prime =n_\phi^{\text{min}}$ in Eq.~\ref{eq:nli}, and we will make a comment about the general case.


Near the end of reheating, the change in the temperature is small such that the temperature as a function of time may be taken as a constant. Hence, the rates $\Gamma_S$, ${\Gamma_\text{1,2 LNV}}_a$ and $\Gamma_{th}$ and the efficiency factor $\epsilon_\text{1,2 a}$ are constant as well, and the evolution equation for the radiation energy density is decoupled from those for the out-of-equilibrium lepton number density and for the net lepton number density. Furthermore, the change with time in the scale factor $a(t)$ is slower than that in the exponential factors such as $e^{-\Gamma_D t}$ 
and $e^{-\Gamma_{th} t}$, since the former is in power law $a(t)\sim t^{\frac{2}{3(1+w)}}$, where $w$ is the equation of state of the system. We therefore temporarily ignore the redshift in $n_\phi^{\rm min}$ ($\propto \rho_\phi(t) \propto e^{-\Gamma_Dt} a(t)^{-3}$) when solving the Boltzmann equations. From Eq.~\ref{eq:nli}, we can easily find
\begin{eqnarray}\label{sol:nli}
n_{l_a}\approx \frac{\Gamma_D}{\Gamma_{th}}\, \mathcal{B}_a\,n_\phi^\text{min}
\end{eqnarray}
where we have used $\Gamma_{th}\gg \Gamma_D$.
Substituting Eq.~\ref{sol:nli} to Eq.~\ref{eq:nL} we obtain
\begin{eqnarray}\label{sol:nL}
n_L\sim \sum_a \mathcal{B}_a\left[4\, \epsilon_{1\, a} \frac{{\Gamma_\text{1 LNV}}_a}{\Gamma_{th}}+2\, \epsilon_{2\, a}\frac{{\Gamma_\text{2 LNV}}_a}{\Gamma_{th}}\right]n_\phi^\text{min}\,
\end{eqnarray}
In the conservative case where the out-of-equilibrium leptons are right after being produced by relaxion decay with an energy of $p^0\to m_\phi/2\ll T$, the first term that we adopted in our manuscript would dominates in Eq.~\ref{sol:nL}. Whereas the second term in Eq.~\ref{sol:nL} would make a similar-sized contribution when the energy of the non-thermal lepton has a value close to $3T$ via upscattering by the thermal bath before reaching a thermal distribution. However, this would not change our main result.

We see from Eq.~\ref{sol:nL} that 
the evolution of the lepton asymmetry mainly arises
due to $\rho_\phi(t)$ in $n_\phi^\text{min}$:
\begin{eqnarray}
n_L(t)\propto 
\rho_\phi(t)\propto e^{-\Gamma_Dt}a(t)^{-3}\propto e^{-\Gamma_Dt}(\Gamma_D t)^{-3/2},\,
\end{eqnarray}
where we have used that the time scale for this process roughly is $t\sim \Gamma_D^{-1}$. At the end of leptogenesis, without entropy injection from the hidden sector, the exponential factor in the evolution only generates a suppression factor of $\mathcal O(1)$. The redshift in $n_L$ and $s$ both scales as $a^{-3}$ and thus does not change the ratio $n_L/s$. For the more general case where the out-of-equilibrium leptons have sources other than the perturbed decaying relaxion, we expect the $n_\phi^\text{min}$ factor in Eq.~\ref{sol:nL} to be replaced by $n_\phi^\prime$.


\end{document}